# DBG2OLC: Efficient Assembly of Large Genomes Using Long Erroneous Reads of the Third Generation Sequencing Technologies

Chengxi Ye[1§], Christopher M. Hill[1], Shigang Wu[2], Jue Ruan[2], Zhanshan (Sam) Ma[3§]

[1]Department of Computer Science, Institute for Advanced Computer Studies, University of Maryland, College Park, MD 20742, USA.

[2]Agricultural Genome Institute, Chinese Academy of Agricultural Sciences, No.7 Pengfei Road, Dapeng New District, Shenzhen, Guangdong 518120, China.

[3]Computational Biology and Medical Ecology Lab, State Key Laboratory of Genetic Resources and Evolution, Kunming Institute of Zoology, Chinese Academy of Sciences, Kunming, 650223 China.

Correspondence email addresses: Chengxi Ye: cxy@umd.edu, Sam Ma: samma@uidaho.edu

## Abstract

The highly anticipated transition from next generation sequencing (NGS) to third generation sequencing (3GS) has been difficult primarily due to high error rates and excessive sequencing cost. The high error rates make the assembly of long erroneous reads of large genomes challenging because existing software solutions are often overwhelmed by error correction tasks. Here we report a hybrid assembly approach that simultaneously utilizes NGS and 3GS data to address both issues. We gain advantages from three general and basic design principles: (*i*) Compact representation of the long reads lead to efficient alignments. (*ii*) Base-level errors can be skipped; structural errors need to be detected and corrected. (*iii*) Structurally correct 3GS reads are assembled and polished. In our implementation, preassembled NGS contigs are used to derive the compact representation of the long reads, which established an algorithmic conversion from a *de Bruijn* graph to an overlap graph, the two major assembly paradigms. Moreover, since NGS and 3GS data can compensate each other, our hybrid assembly approach reduces both of their sequencing requirements. Experiments show that our software is able to assemble mammalian-sized genomes orders of magnitude more efficiently in time than existing methods, while saving about half of the sequencing cost.

**Availability:** The source code and a compiled version of DBG2OLC is available in the following site: https://github.com/yechengxi/DBG2OLC

## Introduction

The Human Genome Project (HGP), which is perhaps the largest biomedical research project humans have ever undertaken, is responsible for greatly accelerating the advancement of DNA sequencing technologies[1]. Three generations of DNA sequencing technologies have been developed in the last three decades, and we are at the crossroads of the second and third generation of the sequencing technologies. The third generation sequencing (3GS) technology promises to significantly improve assembly quality and expand its applications in biomedical research and biotechnology development. However, lack of efficient and effective genome assembly algorithms has arguably been the biggest roadblock in preventing the widespread adoption of 3GS technologies. Packages that are capable of assembling highly desirable 3GS long reads (averaging up to 5-20kb per run at this time) are usually laden with excessively high error rates. At present, the reported error rates are ~15% with PacBio sequencing [2], and can be as high as ~40% with Oxford Nanopore sequencing [3]. These high error rates make the assembly of 3GS sequences seem disproportionally complex and expensive when compared to the assembly of prevailing NGS sequences. As a comparison, the whole genome assembly of a human genome using 3GS data was first reported to have taken half a million CPU hours[4] compared to ~24 hours with Illumina NGS sequencing data[5]. Consequently, in practice, many applications of 3GS technology have been limited to re-sequencing bacteria and other small genomes[6]. In spite of significant efforts and advances in software technology for 3GS assembly[2,7-17], an efficient and effective genome assembler is still highly anticipated to facilitate the technology upgrade from prevailing NGS to upcoming 3GS technology. Another major issue is that the higher sequencing cost of 3GS technology, while decreasing with time, is still at least an order of magnitude more expansive than the popular Illumina NGS sequencing at the time of this work.

While the evolution of genome assembly software solutions has been influenced by multiple factors, the most significant one has been the length of the sequences[18]. Although increasing sequence lengths may simplify the assembly graph[6], the read length also has critical impacts on the computational complexities of genome assembly. Computational biologists have historically formulated the genome assembly problem as a graph traversal problem[18-20], *i.e.*, searching for a most likely genome sequence from the *overlap graph* of the sequence reads in the case of the first generation sequencing technology. The *string graph* and the *best overlap graph* are specific forms of the *Overlap-Layout-Consensus* (OLC) paradigm that are more efficient by simplifying the global overlap graph[19,21,22]. The read-based algorithms, aiming to chain the sequencing reads in the most effective way, are computationally expensive because pair-wise alignment of the sequences is required to construct the overlap graph. This issue was tolerable for the relatively low amount of sequences produced from the low-throughput first generation sequencing technologies, but quickly became overwhelming with the enormous amount of short reads produced by high-throughput NGS data. The strategy of chopping the sequencing reads into shorter and overlapping *k*-grams (so-termed *k*-mers) and building links between the *k*-mers, was developed in the *de Bruijn* graph (DBG) framework to simplify NGS assembly. Assembly results are extracted from the linear (unbranched) regions of the *k*-mer graph in this approach[20].

The overlap graph model or the OLC-based software packages, such as Celera Assembler[1], AMOS[23] and ARACHNE[24], originally used for assembling the first generation sequence data, were also adopted for the NGS assembly before DBG based approaches became the *de facto* standard. Newer 3GS technologies, including single molecule real time sequencing (SMRT) and Oxford Nanopore sequencing, can produce much longer reads than NGS does. The longer reads from 3GS technology make the OLC approaches, which were originally used in the first generation genome assembly, attractive again. Nevertheless, the high error rates of current 3GS technologies render the existing OLC-based assemblers developed for relatively accurate sequences unscalable. Similarly, the error prone long reads make the DBG full of branches and therefore unsuitable for 3GS assembly as well. Faced with these challenges, the developers of 3GS technology have resorted to using error correction techniques[2,7,9,10,13,17] to create high quality long reads and to reusing the algorithms originally developed for the first generation sequence assembly. However, error corrections for these long reads require excessively high computational resources, even for small microbial genomes. Moreover, the high sequencing depth (usually 50x-100x) required by existing 3GS genome assemblers increases sequencing cost significantly, especially for large genomes. Both the issues have put 3GS technology at a severe disadvantage when competing against widely used NGS technology. In this article, we introduce algorithmic techniques that effectively resolve the issues discussed earlier. But first, we present a brief account of the existing genome assembly software technologies to put our contribution in proper context.

Researchers began with scaffolding approaches such as AHA[16], PBJelly[15] and SSPACE-LongRead[11] to patch the gaps between high quality assembly regions, *i.e.*, first build a scaffold by aligning reads to the contigs and then use reads that span multiple contigs as links to build a scaffold graph. In ALLPATHS-LG[14] and *Cerulean* [12], the long reads were used to guide a graph search, a best path in the *de Bruijn* graph was searched to bridge the gaps between large contigs. While these software packages have indeed achieved important advances for 3GS genome assembly, resolving intricate ambiguities is inherently difficult and can easily lead to structural errors. Furthermore, the underlying graph search algorithm usually has *exponential* complexity with respect to the search depth and scales badly; highly repeating regions (such as long repeats of simple sequences) will lead to large search depths and are not resolvable at all. In addition, the more powerful read overlap graph structure (of the long reads) was not fully explored in all these approaches. Often these algorithms rely on heuristics such as contig lengths and require iterations[12,14]. To circumvent these important issues associated with the hybrid approach, a Hierarchical Genome-Assembly Process (HGAP)[13] was developed using a non-hybrid strategy to assemble PacBio SMRT sequencing data, which does not use the NGS short reads. HGAP contains a consensus algorithm that creates long and highly accurate overlapping sequences by correcting errors on the longest reads using shorter reads from the same library. This correction approach was proposed earlier in the hybrid setting and is used in many widely used assembly pipelines[2,9,10,17]. Nonetheless, this non-hybrid, hierarchical assembly approach requires relatively high sequencing coverage (50x-100x) and substantial error correction time to obtain satisfactory results. It is noteworthy that most of the algorithms we reviewed here were originally designed for bacterial-sized genomes. Though recent advancements in aligning erroneous long reads[6,25] have also shortened the computational time of 3GS assembly, running these programs on large genomes especially mammalian-sized genomes usually requires multitudinous CPU hours (sometimes up to $10^5$ or $10^6$ CPU hours) that requires the power of super computers, and well beyond the reach of a typical workstation.

In this study, we design algorithms to enable efficient assembly of large mammalian sized genomes. We observe that per-base error-correction of each long erroneous reads and their pair-wise alignment takes a significantly large portion of time in existing pipelines. But neither of these is necessary at an initial assembly stage. If all sequencing reads are structurally correct, one can produce a structurally correct draft genome and improve the base-level accuracy in the final stage, as was originally done in the OLC approach. Taking advantage of this observation, we develop a base-level correction-free assembly pipeline by directly analyzing and exploiting overlap information in the long reads. Unlike previous approaches, we use the NGS assembly to lower the computational burden of aligning 3GS sequences rather than just polishing 3GS data. This allows us to take advantage of the cheap and easily accessible NGS reads, while avoiding the issues associated with existing hybrid approaches mentioned previously. Meanwhile, since NGS and 3GS are independent of each other, the sequencing gaps in one type of data may be covered by the data from the other. The utilization of NGS data also lowers the sequencing depth required by using 3GS only (*i.e.*, the non-hybrid assembly), and the net result is reduced sequencing cost. Hence, we get the best of both worlds of hybrid and non-hybrid assembly approaches. Specifically, we map the DBG contigs from NGS data to the 3GS long reads to create *anchors* for the long reads. Each long read is (lossy) compressed into a list of NGS *contig identifiers*. Because the compressed reads are often orders of magnitude shorter than the original reads, finding candidate overlaps between them becomes a simple bookkeeping problem and the approximate *alignments* and *overlaps* can be calculated cheaply with the help of the *contig indentifiers*. An overlap graph is constructed by chaining the best overlapped-reads in the compressed domain. The linear unbranched regions of the overlap graph are extracted and uncompressed to construct the draft assembly. Finally, we polish the draft assembly at the base-level with a consensus module to finish the assembly. Overall, compared with the existing approaches, our algorithm offers an efficient algorithmic solution for assembling large genomes with 3GS data in terms of computational resource (time and memory) consumptions and sequencing coverage requirement while also being robust to sequencing errors. Furthermore, our pipeline utilizes the reads overlap information directly and provides an efficient solution to the traditional read threading problem, which is valuable both theoretically and practically even for the NGS assembly[20,26].

## Methods and Implementations

Our algorithm starts with linear unambiguous regions of a *de Bruijn* graph (DBG), and ends up with linear unambiguous regions in an overlap graph (used in the Overlap-Layout-Consensus framework). Due to this property, we dub our software DBG2OLC. The whole algorithm consists of the following five procedures, and we implement them as a pipeline in DBG2OLC. Each piece of the pipeline can be carried out efficiently.

(1) Construct a *de Bruijn* graph (DBG) and output contigs from highly accurate NGS short reads.
(2) Map the contigs to each long read to anchor the long reads. The long reads are compressed into a list of *contig identifiers* to reduce the cost of processing the long reads (Fig. 1a).
(3) Use multiple sequence alignment to clean the compressed reads, and remove reads with structural errors (or so-called chimeras) (Fig. 2).
(4) Construct a best overlap graph using the cleaned compressed long reads (Fig. 1b).
(5) Uncompress and chain together the long reads (Fig. 1c), and resort to a consensus algorithm to convert them into DNA sequences (Fig. 1d).

Details for procedure (2)-(5) are explained below. The explanation of procedure (1) can be found in our previous *SparseAssembler* for NGS technology [5] and is omitted here.

**Reads Compression**

We use a simple *k*-mer index technique to index each DBG contig and map the pre-assembled NGS contigs back to the raw sequencing reads as anchors. The *k*-mers that appear in multiple contigs are excluded in our analysis to avoid ambiguity. Empirically for PacBio reads, we found that using k=17 were adequate for all our experiments. For each 3GS long read, we report the matching contig identifiers as an ordered list. A contig identifier is reported if the number of uniquely matching *k*-mers in that contig is above a threshold, which is adaptively determined based on the contig length. We set this threshold in the range of (0.001~0.02)*Contig_Length. This easily tuneable threshold parameter allows the user to find a balance between sensitivity and specificity. With low coverage datasets, this parameter is set lower to achieve better sensitivity; otherwise it is set to higher to enforce better accuracy. In all our experiments, the contigs are generated with our previous *SparseAssembler*[5].

After this procedure, each read is converted into an ordered list of *contig identifiers*. An example of such a list is {Contig_a, Contig_b}, where Contig_a and Contig_b are identifiers of two different contigs. We also record the orientations of these contigs in the mapping. This compact representation is a lossy compression of the original long reads. We term the converted reads as compressed reads in this work. A compressed read is considered to be equivalent to its reverse complement, and the same compressed reads are then collapsed. Since a *de Bruijn* graph can efficiently partition the genome into chunks of bases as contigs, this lossy compression leads to orders of magnitude reduction in data size. Moreover, the compact representation can span through small regions with low or even no NGS coverage; these important *gap regions* in NGS assembly can be covered by 3GS data. Likewise, small 3GS sequencing gaps may be covered by NGS contigs. These sequencing gaps will be bridged in the final stage. Similarity detection between these compressed reads becomes a simpler bookkeeping problem with the identifiers and can be done quickly with low memory. To demonstrate the effectiveness of this strategy, we ran it over five datasets including genomes of different sizes and different sequencing technologies (Table 1, resources can be found in the supplementary materials). The compression usually leads to three factors of reduction in read length with 3GS.

Table 1. The demonstration of the compression ratio on various datasets.

| Datasets | Sequencing Technology | Average Raw Read Length | NGS Contig N50 (DBG $k$-mer size) | Average Compressed Read Length | Compression Ratio |
|---|---|---|---|---|---|
| *S. cer w303* | PacBio | 4,734 | 31,233 ($k$=51) | 7 | 1:676 |
| *A. thaliana ler-0* | PacBio | 5,614 | 2,264 ($k$=51) | 8 | 1:702 |
| *H. sapiens* | PacBio | 14,519 | 3,115 ($k$=51) | 11 | 1:1320 |
| *E.coli K12* | Oxford Nanopore | 6,597 | 3,303 ($k$=21) | 4 | 1:1649 |
| *E.coli K12* | Illumina Miseq | 150 | 3,303 ($k$=21) | 2 | 1:75 |

**Ultra-fast Pair-wise Alignments**

Most existing algorithms rely on sensitive algorithms[27,28] to align reads to other reads or assemblies. In our approach, since the compressed reads are usually much shorter than the original reads, alignments of these compressed reads can be calculated far more efficiently. We adopt a simple bookkeeping strategy and use the contig identifiers to build an inverted-index. Each identifier points to a set of compressed reads that contain this identifier. This inverted-index helps us to quickly select the potentially overlapped reads based on shared contig identifiers. Alignments are calculated only with these candidate compressed reads. The alignment score is calculated using the Smith-Waterman algorithm[29]; the contig identifiers that can be matched are positively scored while the mismatched contig identifiers are penalized. Scores for match/mismatch are calculated based on the involved contig lengths or the number of matching *k*-mers in the previous step. With the compressed reads, our algorithm can finish pair-wise alignments in a small amount of time.

As discussed previously, state-of-the-art assembly pipelines usually resort to costly base-level error correction algorithms to correct each individual read[2,7,8,10,13], then them feed into an existing assembler. However, an *important finding* of this work is that per-base accuracy may not be a major roadblock for assembly contiguity. Rather, the *chimeras* or *structural sequencing errors* are the major "hot spots" worth putting major effort into. Without cleaning these chimeras, the overlap relations include many falsely generated reads and will lead to a tangled overlap graph. To resolve this issue, we compute multiple sequence alignments (MSA) by aligning each compressed read with all other candidate compressed reads. With MSA we can detect the chimeric reads and the spurious contig identifiers in each read (Fig 2). Both of these errors are cleaned up. The major side effect of this correction is a slightly increased requirement of the 3GS data coverage so that each compressed read can be confirmed by at least another one. The remaining minor errors (mostly false negatives) in the cleaned compressed reads will be tolerated by the alignment algorithm. In our experiments, we noticed that the algorithm is accurate enough to find high quality overlaps and can be used for constructing draft genomes as assembly backbones.

**Read Overlap Graph**

Compared with most hybrid approaches that used long reads to link together the short read contigs, our approach takes the unorthodox way – we use the short read contigs to help link together the long reads. We construct a best overlap graph[21] using the above-described alignment algorithm with the compressed reads. In the best overlap graph, each node represents a compressed read. For each node, the best overlapped nodes (one before and the other after) are found based on the overlap score, and the links between these nodes are recorded. The overlap graph is calculated in two rounds (Fig. 1b). In round *1*, all the contained nodes (with respect to other nodes) are filtered off. For example, {Contig_a, Contig_b} is removed if {Contig_a, Contig_b,

Contig_c} is present. With this strategy, alignments with repeating and contained nodes are avoided. In round *2*, all suffix-prefix overlaps among the remaining nodes are detected with the alignment algorithm. Nodes are chained one to another in both directions and in the best overlapped fashion. Graph simplification is applied to remove tiny tips and merge bubbles in the best overlap graph. Truly unresolvable repeats result in branches in the graph[21] and will be kept as the assembly breakpoints.

Note that constructing the overlap graph with the compressed reads offers us several major benefits. (1) Long read information is sufficiently utilized. (2) The costly long read alignments are accelerated with the easily available NGS contigs. (3) The expensive graph search algorithms (with exponential complexity to the search depth) often used for graph resolving in many existing genome assembly programs are no long needed in our software.

**Consensus**

It is noteworthy that only in this final stage that the compressed reads are converted back to the raw nucleotide reads for polishing purpose. Linear unbranched regions of the best overlap graph encode the unambiguously assembled sequences. Uncompressed long reads that lie in these regions are laid out in the best-overlapped fashion and patched one after another (Fig. 1c). NGS contigs are included when there is a gap in the 3GS data. Reads that are related to each backbone are collected based on the contig identifiers. A consensus module is finally called to align these reads to each backbone and calculate the polished assembly (Fig. 1d). To polish the 3GS assembly backbone, we use an efficient consensus module *Sparc*[30]. *Sparc* builds an efficient sparse *k*-mer graph structure[5] using a collection of sequences from the same genomic region. The heaviest path approximates the most likely genome sequence (consensus) and is found in a sparsity-induced reweighted graph.

## Results

We conducted a comprehensive comparison on a small yeast genome (12Mbp) dataset to provide a scope of the performance of each software program we compared in this study. Since most other programs do not scale linearly with the data scale and require thousands of hours per-run on genomes larger than 100Mbp, the readers are encouraged to read through their original publications for the performance results of those programs.

As a side note, the advent of 3GS long reads has raised the bar to a higher level compared with previous sequencing techniques: existing reference genomes usually contain a large number of structural errors and/or variations that can surpass the number of assembly errors using the long reads. In most cases we select assemblies by other assemblers with more coverage (~100x) as references. If high quality reference genome is available, thorough evaluations of our algorithm show that DBG2OLC can provide high quality results with fewer structural errors and comparable per-base accuracy. This has been recently demonstrated in the case study of *D. melanogaster* genome by Chakraborty et al. (2015)[31] who compared our pre-released software with other premier programs for 3GS data. In this paper, we demonstrate results on some other well-studied species and use existing high coverage assemblies as quality checks. On medium to large genomes, DBG2OLC can produce comparably good results with one to two orders of magnitude less time and memory usages than most existing pipelines. A draft assembly (without polishing) of a 3 Gbp *H. sapiens* can be finished in 3 CPU days with our pipeline, utilizing 30x 3GS and 50x NGS data. This computational time is roughly comparable to many existing NGS assemblers. The time consumption of each step running different genomes can be found in Table 2.

We compared our algorithm results with Celera Assembler (CA, version 8.3rc2), PacBioToCA (in CA8.3rc2)[2], ECTools[9], MHAP (in CA8.3rc2)[7], HGAP (in SMRT Analysis v2.3.0)[13] and Falcon assembler (v0.3.0), which are well recognized as the best-performed genome assemblers for 3GS technologies. Data from PacBio SMRT RS-II (the currently leading platform of 3GS technology) was used to perform the comparative experiments (50x Illumina MiSeq reads were additionally used for PacBioToCA and DBG2OLC, the two hybrid methods). The experiments are run on a server with eight Intel Xeon E7-8857 v2 CPUs (each has 12 cores) and 2TB memory. For all DBG2OLC experiments in this paper we used SparseAssembler (Ye et al. 2012) to preassemble 50x Illumina short reads into contigs and then to compress the 3GS reads. Similarly, Celera Assembler was used to assemble the same short reads into contigs for ECTools.

Unassembled short reads were fed into PacBioToCA according to its specification. At the time of this work, 50x Illumina reads cost less, and also can be obtained more easily, than 1x 3GS reads. Celera Assembler could be run with uncorrected reads on small datasets, so we run it as a baseline.

Table 2. Computation time of each procedure

| Species | Long Read Source | Short Read Assembly (CPU hr) | Compression (CPU hr) | Graph Construction (CPU hr) | Consensus (CPU hr) |
|---|---|---|---|---|---|
| *S. cer w303* | 20x PacBio | 0.1 | 0.03 | 0.005 | 2 |
| *A. thaliana ler-0* | 40x PacBio | 1 | 0.6 | 0.2 | 18 |
| *H. sapians* | 30x PacBio | 25 | 37 | 3 | 2000 |
| *E.coli K12* | 30x Nanopore | 0.1 | 0.02 | 0.002 | 2 |

It is noteworthy that in our current implementation, most of the computation time (~90%) is spent on the consensus step, in which Blasr[28] is called to align all raw reads to the assembly backbone. Since the alignments are multi-threaded, the wall time can be much shorter, depending on the available threads. The consensus step is relatively independent in genome assembly and is open to any future improvements and accelerations. The overall computational time of the whole pipeline scales near linearly to the data size which is a highly valuable property to large-scale genome assembly problems. Using 10x-20x PacBio coverage data, we obtained assembly N50s that are significantly (>10x) better than Illumina data alone (Table 1). The datasets, commands and parameters can be found in the supplementary materials. We used QUAST 3.0[32] in its default setting to evaluate the assembly results; these are reported as the NGA50, per-base identity rates and misassembly errors. In analyzing 3GS assembly results, the NGA50 is a measure of the average length of high quality region before reaching a poor quality region in the assembly. The identity rates were calculated by summarizing the single base mismatches and insertion/deletion mismatches. Relocations, inversions and translocations are regarded as the misassembly errors[32]. The alignment dot plots can be found in the supplementary materials. The nearly perfect diagonal dot plots indicate that DBG2OLC can produce structurally correct assemblies from as low as 10x long read data.

For the yeast dataset we picked an assembly from 454 data (NCBI Accession No.: GCA_000292815.1) and the other one generated using MHAP and high coverage data as references. DBG2OLC can take good advantage of different sequencing types and obtain the most contiguous results using 10x-40x data with comparable levels of accuracy (Table 3). Some non-hybrid assemblers are not able to fully assemble the yeast genome with 10x-20x PacBio data. It is also worth mentioning a caveat in many current hybrid error correction approaches. These pipelines use NGS contigs to correct the 3GS reads, which seem to have improved the accuracy of each individual 3GS read. However, the errors in NGS contigs may have corrupted the originally correct 3GS reads and lead to consensus errors in the final assembly, as we notice the identity rates of ECTools assembly is higher when aligned to the 454 reference, contrary to all other pipelines. With high enough coverage (also significantly increased sequencing cost), the 3GS self-correction based assembly methods catch up and can produce better assembly results. Since our pipeline has a major advantage in low coverage data and efficiency, it is expected to scale well to large genomes where low coverage data and computational time becomes major concerns.

Table 3. Assembly performance comparison on the *S. cerevisiae* genome (genome size: 12M bp)

| Cov | Assembler | Time (h) | NG50 | Contigs | NGA50 (454) | Identity (454) | Misass-emblies (454) | NGA50 (PacBio) | Identity (PacBio) | Misass-emblies (PacBio) | Longest | Sum |
|---|---|---|---|---|---|---|---|---|---|---|---|---|
| 10x | MHAP* | - | - | - | - | - | - | - | - | - | - | - |
|  | HGAP* | 36.3 | - | 554 | - | 99.68% | 105 | - | 99.77% | 6 | 36,942 | 1,512,911 |
|  | CA* | 15.1 | 85,728 | 289 | 68,030 | 97.49% | 134 | 81,451 | 97.46% | 13 | 448,177 | 12,285,888 |
|  | PacBioToCA | 173.5 | 19,694 | 898 | 19,378 | 99.88% | 112 | 18,689 | 99.90% | 6 | 221,736 | 10,741,663 |
|  | ECTools | 24.5 | 120,126 | 169 | 98,965 | 99.76% | 324 | 109,640 | 99.73% | 29 | 525,820 | 11,785,741 |
|  | Falcon* | 1.3 | - | 675 | - | 99.23% | 116 | - | 99.28% | 4 | 36,616 | 4,137,485 |
|  | DBG2OLC | 1.7 | 475,890 | 67 | 168,612 | 99.70% | 408 | 355,269 | 99.81% | 46 | 1,174,277 | 11,899,604 |
| 20x | MHAP* | 17.1 | 241,394 | 87 | 155,221 | 99.70% | 508 | 241,260 | 99.75% | 22 | 490,764 | 12,123,145 |
|  | HGAP* | 31.1 | 8,578 | 1,210 | 6,908 | 99.85% | 307 | 7,619 | 99.90% | 20 | 86,998 | 8,624,090 |
|  | CA* | 42.4 | 371,115 | 165 | 201,649 | 98.83% | 284 | 329,930 | 98.82% | 21 | 680,599 | 13,052,212 |
|  | PacBioToCA | 400.9 | 66,974 | 395 | 65,171 | 99.87% | 157 | 65,171 | 99.91% | 7 | 628,280 | 11,487,222 |
|  | ECTools | 34.2 | 176,663 | 172 | 109,931 | 99.77% | 565 | 150,351 | 99.74% | 46 | 624,112 | 12,887,799 |
|  | Falcon* | 3.5 | 110,083 | 180 | 93,385 | 99.38% | 345 | 110,438 | 99.42% | 15 | 281,041 | 10,583,868 |
|  | DBG2OLC | 2.6 | 597,541 | 47 | 172,455 | 99.71% | 440 | 576,287 | 99.88% | 37 | 1,085,773 | 12,476,994 |
| 40x | MHAP* | 36.6 | 614,363 | 65 | 243,012 | 99.91% | 598 | 589,044 | 99.94% | 24 | 1,090,578 | 12,356,826 |
|  | HGAP* | 36.2 | 211,631 | 93 | 198,387 | 99.94% | 528 | 348,754 | 99.99% | 30 | 796,762 | 12,387,287 |
|  | CA* | 115.2 | 365,912 | 114 | 160,867 | 99.66% | 358 | 377,360 | 99.60% | 11 | 769,189 | 15,171,228 |
|  | PacBioToCA | 621.7 | 96,817 | 371 | 96,476 | 99.87% | 178 | 94,480 | 99.91% | 6 | 742,046 | 11,700,172 |
|  | ECTools | 55.8 | 255,956 | 271 | 166,945 | 99.79% | 891 | 214,377 | 99.76% | 64 | 714,196 | 14,481,947 |
|  | Falcon* | 11.2 | 614,509 | 58 | 247,745 | 99.72% | 336 | 555,886 | 99.74% | 10 | 1,069,920 | 12,116,235 |
|  | DBG2OLC | 4.2 | 672,955 | 28 | 238,683 | 99.87% | 431 | 544,679 | 99.90% | 36 | 1,086,380 | 12,149,997 |
| 80x | MHAP* | 13.5 | 751,122 | 43 | 248,079 | 99.91% | 526 | 745,563 | 99.95% | 10 | 1,537,433 | 12,350,704 |
|  | HGAP* | 46.5 | 818,775 | 33 | 248,655 | 99.95% | 534 | 678,552 | 99.99% | 23 | 1,545,906 | 12,621,393 |
|  | CA* | 236.0 | 430,552 | 75 | 201,397 | 99.80% | 319 | 397,774 | 99.74% | 12 | 984,295 | 16,571,250 |
|  | PacBioToCA | 274.3 | 64,967 | 364 | 63,651 | 99.88% | 45 | 62,268 | 99.91% | 10 | 233,799 | 11,651,218 |
|  | ECTools | 100.9 | 247,871 | 382 | 154,348 | 99.79% | 1,470 | 164,839 | 99.76% | 101 | 881,635 | 15,925,328 |
|  | Falcon* | 34.7 | 810,136 | 99 | 247,480 | 99.81% | 437 | 810,134 | 99.82% | 24 | 1,537,463 | 12,681,860 |
|  | DBG2OLC | 8.1 | 678,365 | 29 | 204065 | 99.92% | 426 | 574,476 | 99.95% | 35 | 1,089,897 | 12,209,592 |

*Assemblers that use only 3GS data.

We tested DBG2OLC on other medium to large genomes from PacBio sequencers (Table 4). On the *A. thaliana* genome (120Mbp), the computations with DBG2OLC finish in one hour, after another hour in constructing the NGS contigs. The consensus module takes another 10-20 CPU hours to get the final assembly. The peak memory usage is 6GB. In comparison, existing pipelines can take over one thousand CPU hours with problems of this scale. On a large 54x human (*H. sapiens*) dataset, DBG2OLC is able to reach assembly with high contiguity starting from 10x PacBio data (NG50 433kbp) and DBG contigs generated from 50x Illumina reads (Table S1 in supplementary materials). To reach a better assembly, the longest 30x of the reads in this dataset (mean length 14.5 kbp) are selected (Table S2 in supplementary materials). DBG2OLC occupies 70GB memory to store the 17-mer index, and takes 37 CPU hours to compress and align the 30x longest PacBio reads. The pair-wise alignment takes only 3 hours and takes less than 6 hours even with the full 54x dataset. The final consensus takes roughly 2000 CPU hours. In an initial report by PacBio scientists, the overlapping process took 405,000 CPU hours[4]. Our final assembly quality (N50=6Mbp) is comparable to the state-of-the-art results obtained using orders of magnitude more resources. When evaluating this assembly, QUAST 3.0 can take weeks to finish the full evaluation even on our best workstation. We therefore only align our assembly to the longest 500Mbp assembly generated by the Pacific Biosciences, and report the NGA50 and identity rate in this portion.

DBG2OLC was also tested on an Oxford Nanopore MinION sequencing dataset (Table 4). According to initial studies, this type of data has even higher (up to ~40%) error rates[3] compared to PacBio sequencing. But we find DBG2OLC still successfully assembled the *E. coli* into one single contig. The polished assembly has an error rate of 0.23%. The dot plot of the alignment of the assembly to the reference can be found in Fig. 13 of the supplementary materials.

Table 4. DBG2OLC assembly performance comparison on various genomes

| Genome | Size | Coverage | NG50 | Contigs | NGA50 | Identity | Misassemblies | Longest | Sum |
|---|---|---|---|---|---|---|---|---|---|
| *A. thaliana* | 120Mbp | 10x PacBio | 405,464 | 881 | 258,924 | 99.77% | 704 | 1,549,329 | 119Mb |
| | | 20x PacBio | 2,431,755 | 306 | 926,138 | 99.90% | 117 | 6,015,430 | 120Mb |
| | | 40x PacBio | 3,601,597 | 243 | 1,605,981 | 99.93% | 131 | 15,473,059 | 129Mb |
| *H. sapiens* | 3.0Gbp | 10x PacBio | 432,739 | 16,689 | 347,104 | 99.56% | - | 3,507,306 | 2.97G |
| | | 20x PacBio | 1,886,756 | 9,757 | 1,416,766 | 99.82% | - | 14,597,500 | 3.13Gb |
| | | 30x longest PacBio | 6,085,133 | 13,095 | 4,124,714 | 99.85% | - | 23,825,526 | 3.21Gb |
| *E. coli* | 4.6Mbp | 30x Nanopore | 4,680,635 | 1 | 1,850,974 | 99.77% | 1 | 4,680,635 | 4.7Mb |

Compared with the state-of-the-art assemblers for 3GS technologies, our proposed method requires lower sequencing coverage and minimum memory, provides high contiguity, and is orders of magnitude faster on large genomes. Its combination of different data types leads to both computation and cost efficiency. These advantages are gained from three general and basic design principles: (*i*) Compact representation of the long reads lead to efficient alignments. (*ii*) Base-level errors can be skipped, but structural errors need to be detected and cleaned. (*iii*) Structurally correct 3GS reads are assembled and polished. DBG2OLC is a specific and simple realization of these principles. Interestingly, this implementation builds a nice connection between the two major assembly frameworks, and even though DBG2OLC is majorly developed for 3GS data, this strategy of compression and converting a *de Bruijn* graph to an overlap graph is general and can be used for popular NGS data. A preliminary show case on a purely NGS dataset can be found in the supplementary materials. The strategy of compressing the long reads and carrying out most expensive computations in the compressed domain strikes a balance between the DBG and OLC frameworks.

## Summary and Discussion

In summary, we have built and validated a new *de novo* assembly pipeline that significantly reduces the computational and sequencing requirements of 3GS assembly. We demonstrate that the erroneous long reads can be directly assembled and can lead to significantly improved assembly without base-level error correction. This strategy, first publicly demonstrated in our pre-released pipeline in 2014, has paved the road for several subsequent development attempts on efficient utilization of 3GS data and promises even more efficient 3GS assemblers. Another major finding in developing DBG2OLC is that 3GS technologies generate chimeric reads, and the problem seems to be severer with the PacBio platform. These structural errors lead to tangles in the assembly graph and greatly hamper the assembly contiguity. The most straightforward way to clean up the chimeric reads resorts to multiple sequence alignment, as implemented in DBG2OLC, which lead to slightly increased coverage requirement. This limitation points directions to future development of sequencing technology and correction algorithms. We conjecture that near perfect assemblies can be reached with even lower coverage if the chimeras/structural errors can be removed.

**Acknowledgements:** We appreciate Prof. Mihai Pop, Prof. James Yorke, Dr. Aleksey Zimin and their groups at the University of Maryland for supports and helpful discussions. We thank Dr. Sergey Koren and Daniel Liang for helping us to improve our manuscript.

# Figures

**Figure 1a**. Map *de Bruijn* graph contigs to the long reads. The long reads are in red, the *de Bruijn* graph contigs are in other colors. Each long read is converted into an ordered list of contigs, termed compressed reads.

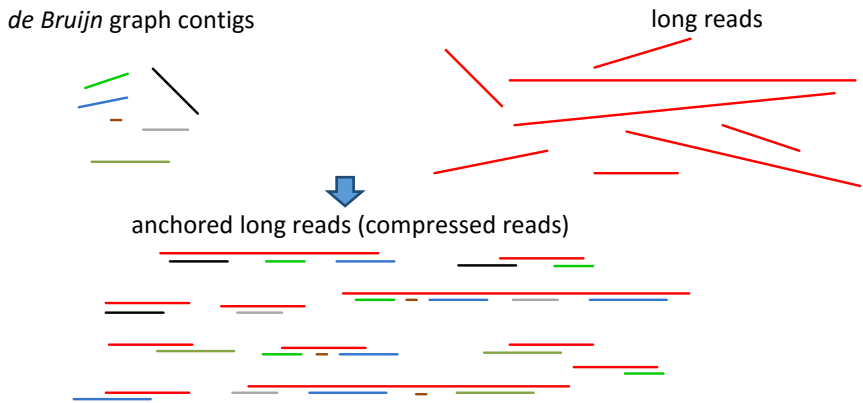

**Figure 1b**. Calculate overlaps between the compressed reads. The alignment is calculated using the anchors. Contained reads are removed and the reads are chained together in the best-overlap fashion.

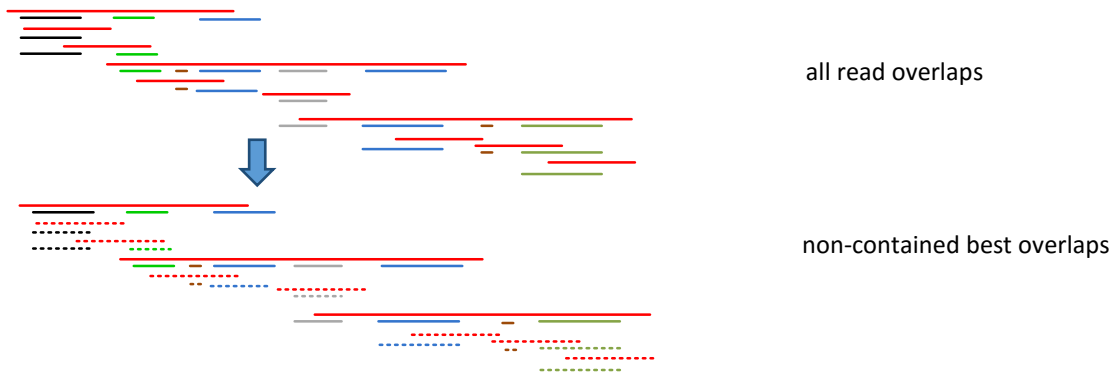

**Figure 1c**. Layout: construct the assembly backbone from the best overlaps.

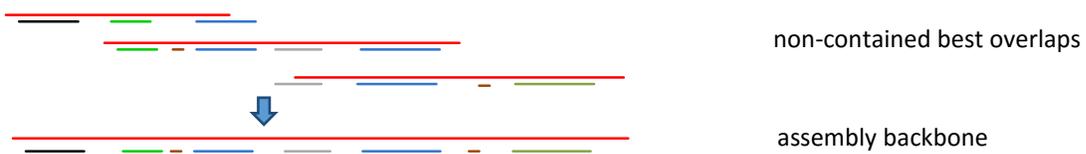

**Figure 1d**. Consensus: align all related reads to the backbone and calculate the most likely sequence as the consensus output.

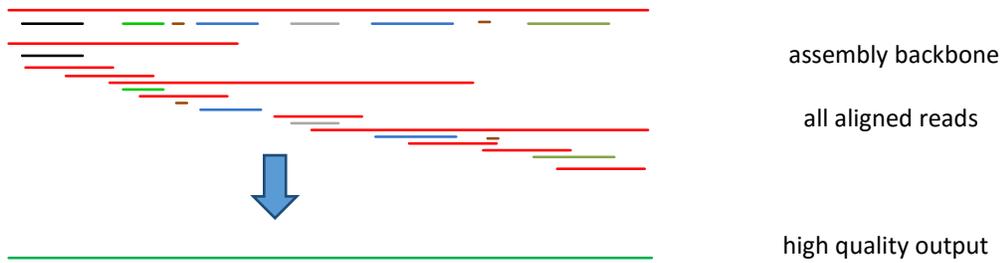

Figure 2. Reads correction by multiple sequence alignment. The left portion shows removing a false positive anchoring contig (brown) that appears only once in the multiple alignment. The right portion shows detection of a chimeric read by aligning it to multiple reads. A breakpoint is detected as all the reads can be aligned with the left portion of the target read are not consistent with all the reads that can be aligned with the right portion of the target read.

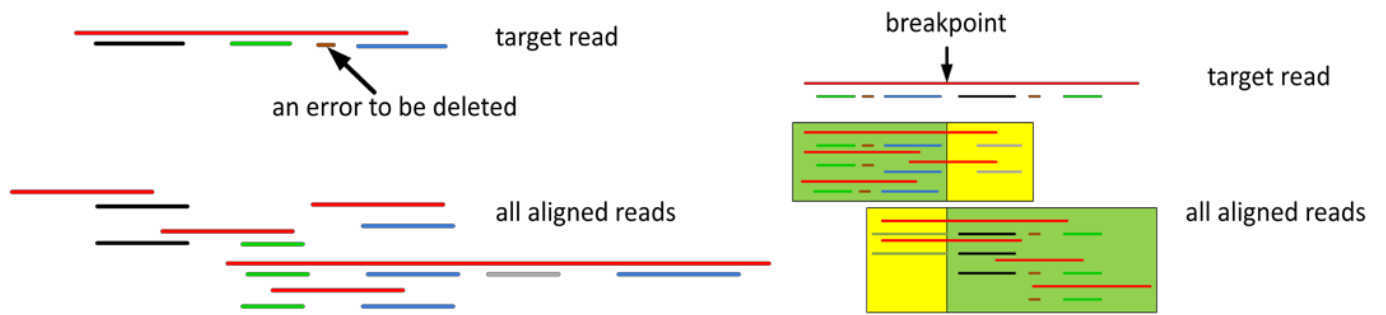

# References

1   Venter, J. C. *et al.* The sequence of the human genome. *Science* **291**, 1304-1351, doi:10.1126/science.1058040 (2001).
2   Koren, S. *et al.* Hybrid error correction and de novo assembly of single-molecule sequencing reads. *Nature biotechnology* **30**, 693-700, doi:10.1038/nbt.2280 (2012).
3   Laver, T. *et al.* Assessing the performance of the Oxford Nanopore Technologies MinION. *Biomolecular Detection and Quantification* **3**, 1-8 (2015).
4   Pacific Biosciences of California, I. *Data Release: ~54x Long-Read Coverage for PacBio-only De Novo Human Genome Assembly*, <http://www.pacb.com/blog/data-release-54x-long-read-coverage-for> (Published: 2014, Date of access: 17/03/2016).
5   Ye, C., Ma, Z. S., Cannon, C. H., Pop, M. & Yu, D. W. Exploiting sparseness in de novo genome assembly. *BMC Bioinformatics* **13 Suppl 6**, S1, doi:10.1186/1471-2105-13-S6-S1 (2012).
6   Koren, S. & Phillippy, A. M. One chromosome, one contig: complete microbial genomes from long-read sequencing and assembly. *Current Opinion in Microbiology* **23**, 110-120, doi:10.1016/j.mib.2014.11.014 (2015).
7   Berlin, K. *et al.* Assembling large genomes with single-molecule sequencing and locality-sensitive hashing. *Nature biotechnology* **33**, 623-630 (2015).
8   Salmela, L. & Rivals, E. LoRDEC: accurate and efficient long read error correction. *Bioinformatics*, doi:10.1093/bioinformatics/btu538 (2014).
9   Lee, H. *et al.* Error correction and assembly complexity of single molecule sequencing reads. *BioRxiv*, 006395, doi:10.1101/006395 (2014).
10  Hackl, T., Hedrich, R., Schultz, J. & Forster, F. proovread: large-scale high-accuracy PacBio correction through iterative short read consensus. *Bioinformatics*, doi:10.1093/bioinformatics/btu392 (2014).
11  Boetzer, M. & Pirovano, W. SSPACE-LongRead: scaffolding bacterial draft genomes using long read sequence information. *BMC Bioinformatics* **15**, 211, doi:10.1186/1471-2105-15-211 (2014).
12  Deshpande, V., Fung, E. K., Pham, S. & Bafna, V. in *Algorithms in Bioinformatics* Vol. 8126 *Lecture Notes in Computer Science* (eds Aaron Darling & Jens Stoye) Ch. 27, 349-363 (Springer Berlin Heidelberg, 2013).
13  Chin, C. S. *et al.* Nonhybrid, finished microbial genome assemblies from long-read SMRT sequencing data. *Nature methods* **10**, 563-569, doi:10.1038/nmeth.2474 (2013).
14  Ribeiro, F. J. *et al.* Finished bacterial genomes from shotgun sequence data. *Genome research* **22**, 2270-2277, doi:10.1101/gr.141515.112 (2012).
15  English, A. C. *et al.* Mind the gap: upgrading genomes with Pacific Biosciences RS long-read sequencing technology. *PloS one* **7**, e47768, doi:10.1371/journal.pone.0047768 (2012).
16  Bashir, A. *et al.* A hybrid approach for the automated finishing of bacterial genomes. *Nature biotechnology* **30**, 701-707, doi:10.1038/nbt.2288 (2012).
17  Au, K. F., Underwood, J. G., Lee, L. & Wong, W. H. Improving PacBio long read accuracy by short read alignment. *PloS one* **7**, e46679, doi:10.1371/journal.pone.0046679 (2012).
18  Nagarajan, N. & Pop, M. Sequence assembly demystified. *Nature reviews. Genetics* **14**, 157-167, doi:10.1038/nrg3367 (2013).

19    Myers, E. W. The fragment assembly string graph. *Bioinformatics* **21 Suppl 2**, ii79-85, doi:10.1093/bioinformatics/bti1114 (2005).
20    Pevzner, P. A., Tang, H. & Waterman, M. S. An Eulerian path approach to DNA fragment assembly. *Proceedings of the National Academy of Sciences* **98**, 9748-9753, doi:10.1073/pnas.171285098 (2001).
21    Miller, J. R. *et al.* Aggressive assembly of pyrosequencing reads with mates. *Bioinformatics* **24**, 2818-2824, doi:10.1093/bioinformatics/btn548 (2008).
22    Simpson, J. T. & Durbin, R. Efficient de novo assembly of large genomes using compressed data structures. *Genome research* **22**, 549-556, doi:10.1101/gr.126953.111 (2012).
23    Treangen, T. J., Sommer, D. D., Angly, F. E., Koren, S. & Pop, M. Next generation sequence assembly with AMOS. *Current Protocols in Bioinformatics*, doi:10.1002/0471250953.bi1108s33 (2011).
24    Batzoglou, S. *et al.* ARACHNE: a whole-genome shotgun assembler. *Genome research* **12**, 177-189, doi:10.1101/gr.208902 (2002).
25    Myers, G. Efficient local alignment discovery amongst noisy long reads. *Algorithms in Bioinformatics*, 52-67, doi:10.1007/978-3-662-44753-6 (2014).
26    Chaisson, M. J., Brinza, D. & Pevzner, P. A. De novo fragment assembly with short mate-paired reads: Does the read length matter? *Genome research* **19**, 336-346, doi:10.1101/gr.079053.108 (2009).
27    Kurtz, S. *et al.* Versatile and open software for comparing large genomes. *Genome Biology* **5**, R12 (2004).
28    Chaisson, M. & Tesler, G. Mapping single molecule sequencing reads using basic local alignment with successive refinement (BLASR): application and theory. *BMC Bioinformatics* **13**, 238 (2012).
29    Smith, T. F. & Waterman, M. S. Identification of Common Molecular Subsequences. *J Mol Biol* **147**, 195-197, doi:10.1016/0022-2836(81)90087-5 (1981).
30    Ye, C. & Ma, Z. S. Sparc: a sparsity-based consensus algorithm for long erroneous sequencing reads. *PeerJ PrePrints 3:e1745* (2015).
31    Chakraborty, M., Baldwin-Brown, J. G., Long, A. D. & Emerson, J. J. A practical guide to de novo genome assembly using long reads. *bioRxiv*, 029306 (2015).
32    Gurevich, A., Saveliev, V., Vyahhi, N. & Tesler, G. QUAST: quality assessment tool for genome assemblies. *Bioinformatics* **29**, 1072-1075, doi:10.1093/bioinformatics/btt086 (2013).

# DBG2OLC: Efficient Assembly of Large Genomes Using Long Erroneous Reads of the Third Generation Sequencing Technologies

Chengxi Ye[1§], Christopher M. Hill[1], Shigang Wu[2], Jue Ruan[2], Zhanshan (Sam) Ma[3§]

[1]Department of Computer Science, Institute for Advanced Computer Studies, University of Maryland, College Park, MD 20742, USA.

[2]Agricultural Genome Institute, Chinese Academy of Agricultural Sciences, No.7 Pengfei Road, Dapeng New District, Shenzhen, Guangdong 518120, China.

[3]Computational Biology and Medical Ecology Lab, State Key Laboratory of Genetic Resources and Evolution, Kunming Institute of Zoology, Chinese Academy of Sciences, Kunming, 650223 China.

Correspondence email addresses: Chengxi Ye: cxy@umd.edu, Sam Ma: samma@uidaho.edu

# Supplementary Materials

## 1. Source Code

The source code of SparseAssembler, DBG2OLC and Sparc can be found here:
https://github.com/yechengxi/SparseAssembler
https://github.com/yechengxi/DBG2OLC
https://github.com/yechengxi/Sparc
To compile, download the code into separate folders and use:

*g++ -O3 -o SparseAssebmler \*.cpp*

*g++ -O3 -o DBG2OLC \*.cpp*

*g++ -O3 -o Sparc \*.cpp*

## 2. Datasets used in the paper

Table S1. Illumina datasets used in the paper

| Datasets | Sequencing Type | Coverage Used | Illumina Data Source |
|---|---|---|---|
| *S. cer* w303 | MiSeq | 50x | http://schatzlab.cshl.edu/data/ectools/ |
| *A. thaliana* ler-0 | MiSeq | 50x | http://schatzlab.cshl.edu/data/ectools/ |
| *H. sapiens* | HiSeq | 50x | Accession No.: SRR1283824 |
| *E.coli* K12 | MiSeq | 50x | Accession No.: SRR826442, SRR826444, SRR826446, SRR826450 |

Table S2. PacBio/Nanopore datasets used in the paper

| Datasets | PacBio Data Source |
|---|---|
| *S. cer* w303 | http://schatzlab.cshl.edu/data/ectools/ |
| *A. thaliana* ler-0 | http://schatzlab.cshl.edu/data/ectools/ |
| *H. sapiens* | http://datasets.pacb.com/2014/Human54x/fast.html |
| *E. coli* K12 | http://gigadb.org/dataset/100102 |

Table S3. Reference genomes used in the paper

| Datasets | Reference Data Source |
|---|---|
| *S. cer* w303 | 1. http://www.cbcb.umd.edu/software/PBcR/mhap/asm/yeast.quiver.all.fasta<br>2. Accession No.: GCA_000292815.1<br>ftp://ftp.ncbi.nlm.nih.gov/genomes/all/GCA_000292815.1_ASM29281v1/GCA_000292815.1_ASM29281v1_genomic.fna.gz |
| *A. thaliana* ler-0 | http://www.cbcb.umd.edu/software/PBcR/mhap/asm/athal.quiver.all.fasta |
| *H. sapiens* | Accession No.: GCA_000772585.3<br>ftp://ftp.ncbi.nlm.nih.gov/genomes/all/GCA_000772585.3_ASM77258v3/GCA_000772585.3_ASM77258v3_genomic.fna.gz |
| *E. coli* K12 | Accession No.: NC_000913<br>ftp://ftp.ncbi.nlm.nih.gov/genomes/all/GCA_000005845.2_ASM584v2/GCA_000005845.2_ASM584v2_genomic.fna.gz |

## 3. Exemplary Assembly Commands

*Step0.* [Optional] Preparations:

We have provided code to help you select a subset of the reads:

https://github.com/yechengxi/AssemblyUtility

The utility functions can be compiled in the same way as the main programs. After compilation, you can use the following command to select a subset of reads from fasta/fastq files. Note that longest 0 is used here, if you set it to 1 it will select the longest reads.

*./SelectLongestReads sum 600000000 longest 0 o Illumina_50x.fastq f Illumina_500bp_2x300_R1.fastq*

*./SelectLongestReads sum 260000000 longest 0 o Pacbio_20x.fasta f Pacbio.fasta*

And you can use the following command to evaluate an assembly.

*./AssemblyStatistics contigs YourAssembly.fasta*

The program will generate two txt files containing essential statistics about your assembly.

*Step1*. Use a DBG-assembler to construct short but accurate contigs. Please make sure they are the **raw** DBG contigs **without** using repeat resolving techniques such as gap closing or scaffolding. Otherwise you may have poor final results due to the errors introduced by the heuristics used in short read assembly pipelines.

SparseAssembler command format:

*./SparseAssembler GS [GENOME_SIZE] NodeCovTh [FALSE_KMER_THRESHOLD] EdgeCovTh [FALSE_EDGE_THRESHOLD] k [KMER_SIZE] g [SKIP_SIZE] f [YOUR_FASTA_OR_FASTQ_FILE1] f [YOUR_FASTA_OR_FASTQ_FILE2] f [YOUR_FASTA_OR_FASTQ_FILE3_ETC]*

A complete example on the *S.cer* w303 dataset:

Download the Illumina reads from

ftp://qb.cshl.edu/schatz/ectools/w303/Illumina_500bp_2x300_R1.fastq.gz

Normally with ~50x coverage, NodeCovTh 1 EdgeCovTh 0 can produce good results.

*./SparseAssembler LD 0 k 51 g 15 NodeCovTh 1 EdgeCovTh 0 GS 12000000 f ../Illumina_data/Illumina_50x.fastq*

In this test run, the N50 is 29 kbp. As we have selected the beginning part of the sequencing file, which usually is of lower quality, the next step may help to improve the assembly quality.

[Miscellaneous]

For other more complex genomes or a different coverage, the first run may not generate the best result. The previous computations can be loaded and two parameters can be fine-tuned to construct a cleaner de Bruijn/ k-mer graph:

*./SparseAssembler **LD 1 NodeCovTh 2 EdgeCovTh 1** k 51 g 15 GS 12000000 f ../Illumina_data/Illumina_50x.fastq*

The N50 is improved to 32kbp in my run.

The output Contigs.txt will be used by DBG2OLC.

*Step2.* Overlap and layout. Feed DBG2OLC with the contig file in fasta format from the previous step (Contigs.txt in this example).

Download the PacBio reads from:

ftp://qb.cshl.edu/schatz/ectools/w303/Pacbio.fasta.gz

The basic command format of DBG2OLC is:

*./DBG2OLC k [KmerSize] AdaptiveTh [THRESH_VALUE1] KmerCovTh [THRESH_VALUE2] MinOverlap [THRESH_VALUE3] Contigs [NGS_CONTIG_FILE] f [LONG_READS.FASTA] RemoveChimera 1*

In the following example, the first 20x PacBio reads are extracted from the abovementioned file and we can assemble with:

*./DBG2OLC k 17 AdaptiveTh 0.0001 KmerCovTh 2 MinOverlap 20 RemoveChimera 1 Contigs Contigs.txt f ../Pacbio_data/Pacbio _20x.fasta*

In our test run, the N50 is 583kbp.

There are three major parameters that affect the assembly quality:

M = matched k-mers between a contig and a long read.

**AdaptiveTh**: adaptive *k*-mer matching threshold. If M < AdaptiveTh* Contig_Length, this contig cannot be used as an anchor to the long read.

**KmerCovTh**: fixed k-mer matching threshold. If M < KmerCovTh, this contig cannot be used as an anchor to the long read.

**MinOverlap**: minimum overlap score between a pair of long reads.

For each pair of long reads, an overlap score is calculated by aligning the compressed reads and score with the matching *k*-mers.

[Miscellaneous]

At this point, the parameters may be fine-tuned to get better performance. As with SparseAssembler, LD 1 can be used to load the compressed reads/anchored reads.

Suggested tuning range is provided here:

For 10x/20x PacBio data: KmerCovTh 2-5, MinOverlap 10-30, AdaptiveTh 0.001~0.01.

For 50x-100x PacBio data: KmerCovTh 2-10, MinOverlap 50-150, AdaptiveTh 0.01-0.02.

Some other less flexible or less important parameters:

k: *k*-mer size, 17 works well.

Contigs: the fasta contigs file from existing assembly.

MinLen: minimum read length.

RemoveChimera: remove chimeric reads in the dataset, suggest 1 if you have >10x coverage.

For high coverage data (100x), there are two other parameters:

ChimeraTh: default: 1, set to 2 if coverage is ~100x.

ContigTh: default: 1, set to 2 if coverage is ~100x.

These two are used in multiple alignment to remove problematic reads and false contig anchors. When we have high coverage, some more stringent conditions shall be applied as with the suggested parameters.

*Step 3.* Call consensus. Install *Blasr* and the consensus module (*Sparc/PBdagcon*). Make sure they are in your path variable.

The input files for consensus are:

(1) backbone_raw.fasta by DBG2OLC

(2) DBG2OLC_Consensus_info.txt by DBG2OLC

(3) DBG contigs (in FASTA format)

(4) PacBio reads (in FASTA format)

You can check the N50 of (1) to see if it meets your standard, otherwise keep tuning and don't proceed.

# this is to concatenate the contigs and the raw reads for consensus

*cat Contigs.txt pb_reads.fasta > ctg_pb.fasta*

# we need to open a lot of files to distribute the above file into lots of smaller files

*ulimit -n unlimited*

#run the consensus scripts

*sh ./split_and_run_sparc.sh backbone_raw.fasta DBG2OLC_Consensus_info.txt ctg_reads.fasta ./consensus_dir 2 >cns_log.txt*

Commands used to assemble other genomes:

The *A. thaliana* Ler-0 dataset:

20x PacBio reads:

*./DBG2OLC KmerCovTh 2 AdaptiveTh 0.005 MinOverlap 20 RemoveChimera 1 Contigs Contigs.txt k 17 f ../PacBio/20x.fasta*

40x PacBio reads:

*./DBG2OLC KmerCovTh 2 AdaptiveTh 0.01 MinOverlap 20 RemoveChimera 1 Contigs Contigs.txt k 17 f ../PacBio/40x.fasta*

The *H. sapiens* dataset:

Longest 30x PacBio reads:

*./DBG2OLC k 17 KmerCovTh 2 MinOverlap 20 AdaptiveTh 0.01 RemoveChimera 1 Contigs Contigs.txt f 30x.fasta >DBG2OLC_LOG.txt*

**Illumina-only Assembly**

With the rapid advancement of sequencing technology, the length of the accurate NGS reads has also become increasingly longer. As a prototype, we demonstrate that the approach can be extended to existing Illumina data. Existing DBG based assembly algorithms resorted to stretches of perfectly matching *k*-mers and are not robust to sequencing errors. Algorithms required growing *k*-mer sizes to find increasingly perfect overlaps and extensive iterations to exploit the long read information. A costly error correction module was critical in finding these perfect overlaps. In contrast, our work can quickly and reliably find the best read overlaps without per-base level correction. Since our approach of utilizing the long read information is certainly not restricted to low quality ones, we compared the performance of our assembler on longer Illumina reads with several popular assemblers, including SGA [1], SOAPdenovo2 [2], SPAdes3 [3]. Interestingly, for the relatively longer short NGS reads generated from the latest NGS technology, traditional *de Bruijn* graph with a fixed *k*-mer size have already exposed its shortage and produces a non-optimal assembly: smaller *k*-mer fails to resolve repetitive regions, while using a large *k* requires high quality and high coverage data. This pair of contradictory requirements makes it hardly possible to obtain the optimal assembly with limited computational resources. Iterative *de Bruijn* graph may partially deal with the problem at the expense of more computational time, as well as more intricate algorithm design and implementations. For example, an error correction procedure (with an exponential-complexity graph search) would be necessary to produce long and correct *k*-mers. SGA utilized the FM-index [4] to find exact matches, which also poses restriction on the quality of the data. In contrast, our algorithm is robust and poses loose restriction on the quality of sequence reads, and hence can find overlaps efficiently and correctly. A dataset of 50X 150bp Illumina Miseq reads of *E. coli* K-12 MG1655 (Accession no: SRA073308) is used here as a test case. SparseAssembler[5] with *k* = 31 is called to assemble the initial contigs and it reaches an N50 of 13.5 kbp. The compressed reads were calculated using the contigs and raw Illumina reads. Our overlap graph construction took only around 1 second on this dataset while the compressing is taking most of the computational time, which is roughly two minutes. The total computational time of SparseAssembler and DBG2OLC is 5 minutes. DBG2OLC exhibits excellent adaptability and overall performance compared with leading assemblers for the new types of the NGS data.

Table S4. Assembly performance comparison on the *E.coli* genome using NGS MiSeq reads (Genome size: 4.6M bp)

| Assembler | Time (hr) | Memory Peak (GB) | NG50 | Identity (%) | NGA50 | Misassemblies | Longest | Sum |
|---|---|---|---|---|---|---|---|---|
| SGA | 1 | 2 | 54,945 | 100.0% | 54,800 | 0 | 185,528 | 4,629,362 |
| SOAPdenovo2 | 0.1 | 0.1 | 54,896 | 100.0% | 52,788 | 0 | 160,084 | 4,600,372 |
| SPAdes3 | 2 | 8 | 112,387 | 100.0% | 107,855 | 0 | 327,031 | 4,659,050 |
| DBG2OLC | 0.1 | 0.1 | 118,892 | 100.0% | 110,457 | 0 | 327,255 | 4,693,366 |

**Commands for Illumina-only assembly:**

The program command is slightly different.

Example command:

*./DBG2OLC LD 0 MinOverlap 70 PathCovTh 3 Contigs Contigs.txt k 31 KmerCovTh 0 f ReadsFile1.fa f ReadsFile2.fq f MoreFiles.xxx*

There are four critical parameters:

k: k-mer length (max size: 31).

**KmerCovTh**: # k-mer matches for a contig to be regarded as a genuine anchor, suggest 0-1.

**MinOverlap**: # 'consistent' k-mers between each pair of reads to be considered to overlap.

**PathCovTh**: the minimum occurrence for a compressed read for a compressed read to be used, suggest 1-3.

Assembly is reported as DBG2OLC_Consensus.fasta.

The command we used for E. coli Illumina Miseq dataset:

*./DBG2OLC k 31 PathCovTh 2 MinLen 50 MinOverlap 31 Contigs Contigs.txt KmerCovTh 0 f Illumina_reads.fasta*

## 4. Dotplots of Alignments to the Reference Genomes

The following commands were used to generate the dot plots:
*nucmer -mumreference -b 200 -g 200 -c 200 –p out Reference.fasta Assembly.fasta*
*mummerplot --png --l --large -f --fat out.delta*

**Figure S1**. The alignment of DBG2OLC assembly of the *S. cer* genome (Y-axis) to the reference created using 454 sequencing (X-axis). 10x PacBio reads and 50x Illumina reads are used.

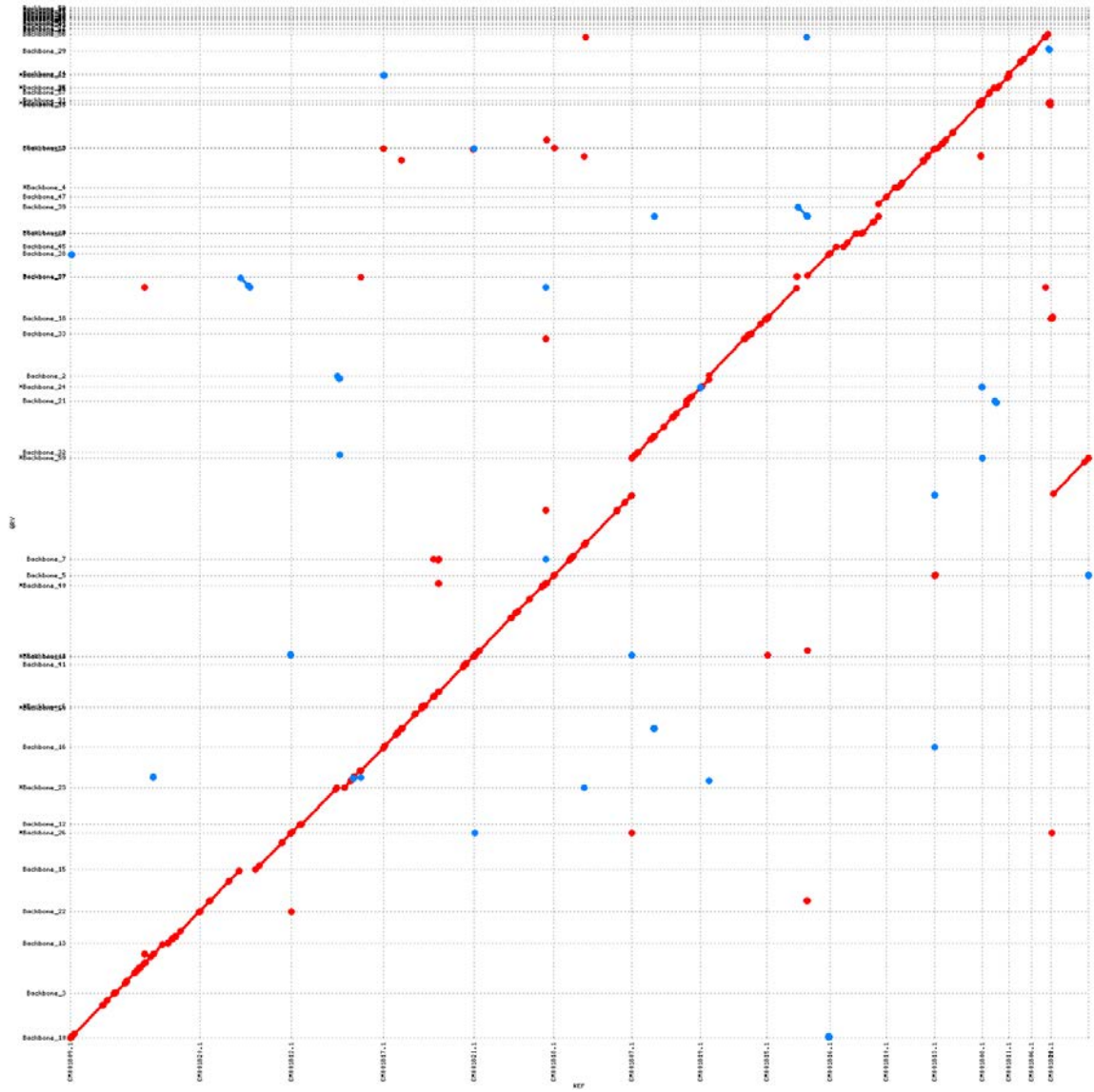

**Figure S2**.The alignment of DBG2OLC assembly of the *S. cer* genome (Y-axis) to the high coverage PacBio assembly (X-axis). 10x PacBio reads and 50x Illumina reads are used.

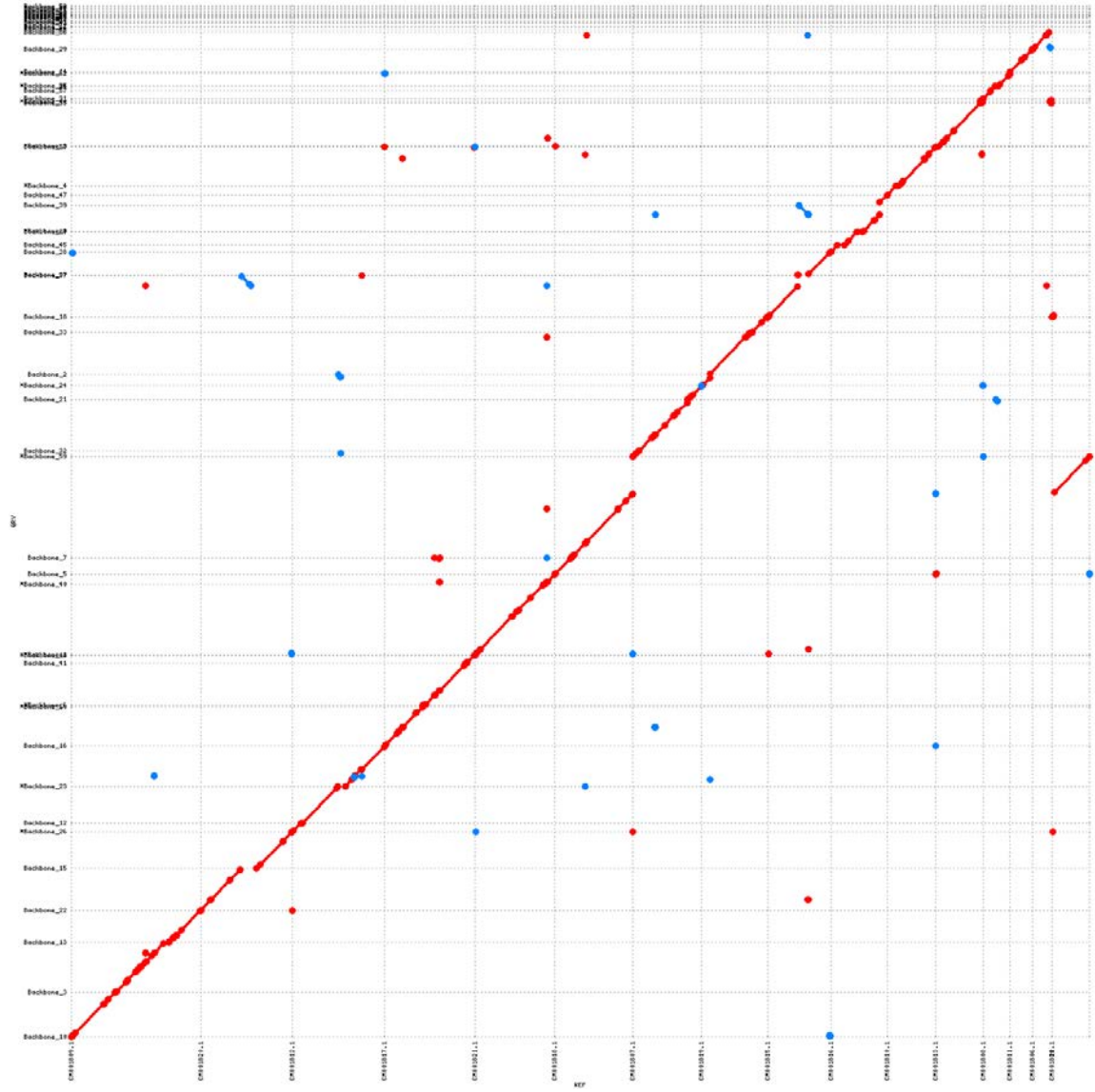

**Figure S3**.The alignment of DBG2OLC assembly of the *S. cer* genome (Y-axis) to the reference created using 454 sequencing (X-axis). 10x PacBio reads and 50x Illumina reads are used.

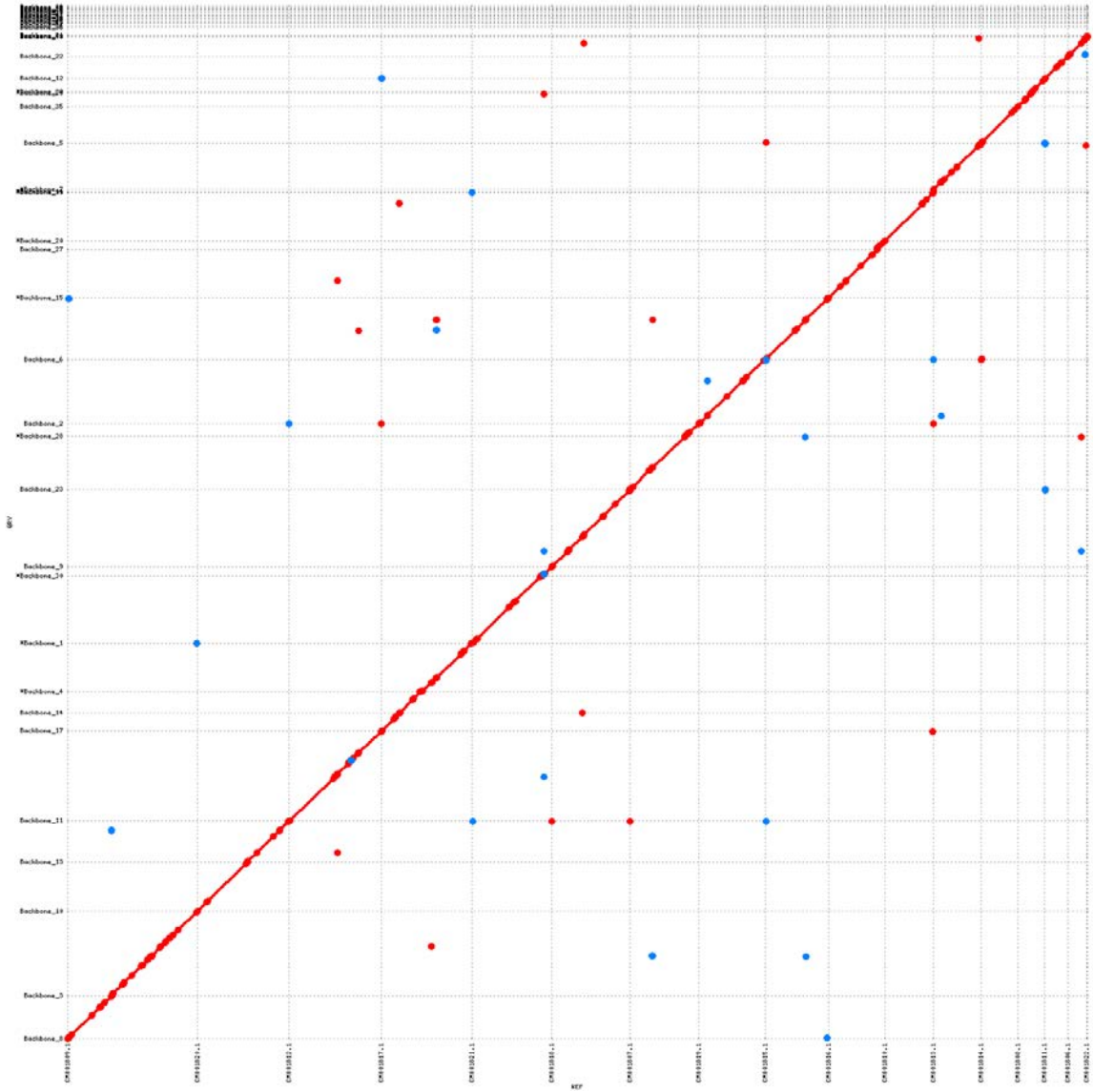

**Figure S4**. The alignment of DBG2OLC assembly of the *S. cer* genome (Y-axis) to the high coverage PacBio assembly (X-axis). 20x PacBio reads and 50x Illumina reads are used.

**Figure S5**.The alignment of DBG2OLC assembly of the *S. cer* genome (Y-axis) to the reference created using 454 sequencing (X-axis). 40x PacBio reads and 50x Illumina reads are used.

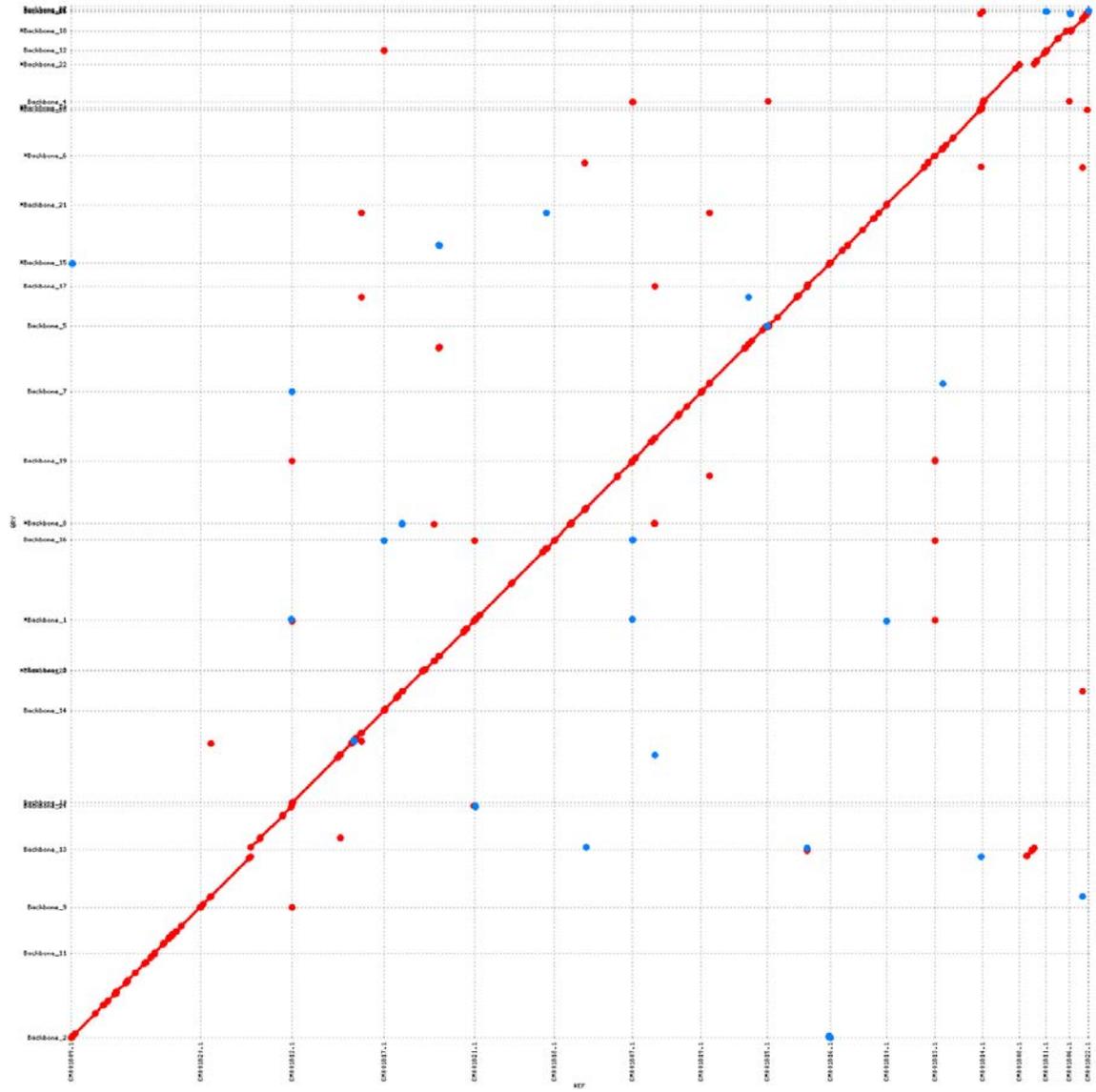

**Figure S6**. The alignment of DBG2OLC assembly of the *S. cer* genome (Y-axis) to the high coverage PacBio assembly (X-axis). 40x PacBio reads and 50x Illumina reads are used.

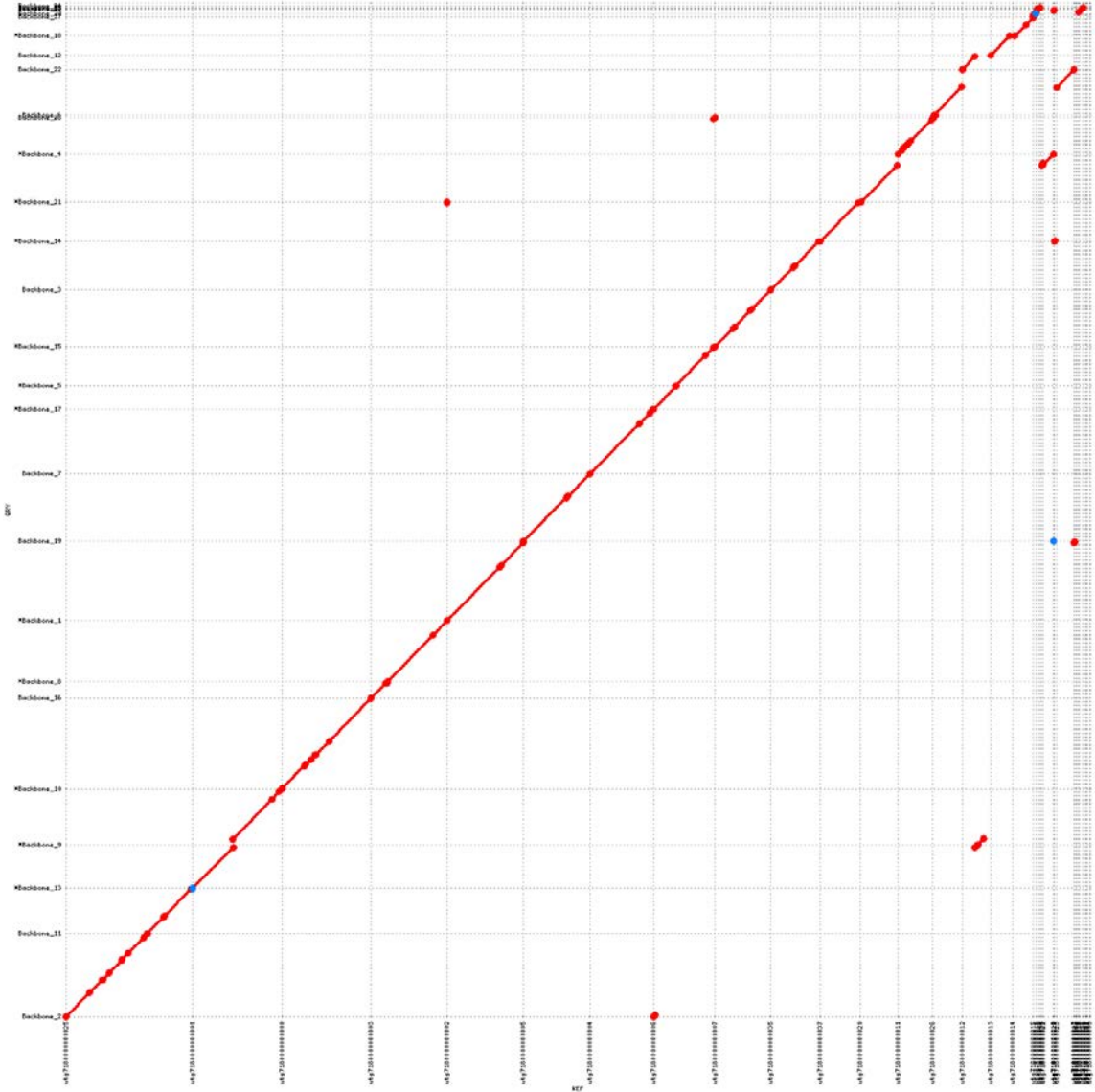

**Figure S7**.The alignment of DBG2OLC assembly of the *S. cer* genome (Y-axis) to the reference created using 454 sequencing (X-axis). 80x PacBio reads and 50x Illumina reads are used.

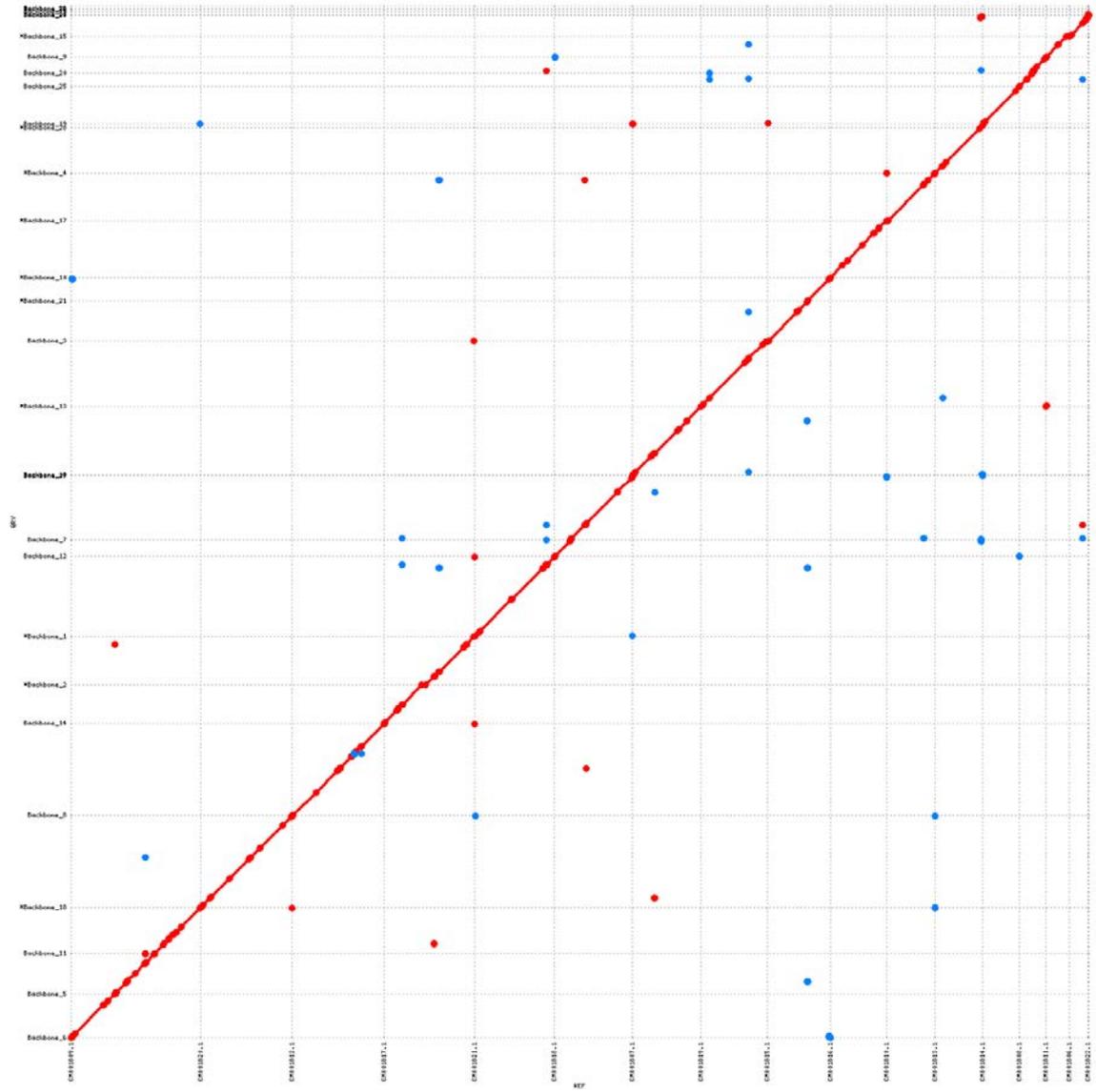

**Figure S8**.The alignment of DBG2OLC assembly of the *S. cer* genome (Y-axis) to the high coverage PacBio assembly (X-axis). 80x PacBio reads and 50x Illumina reads are used.

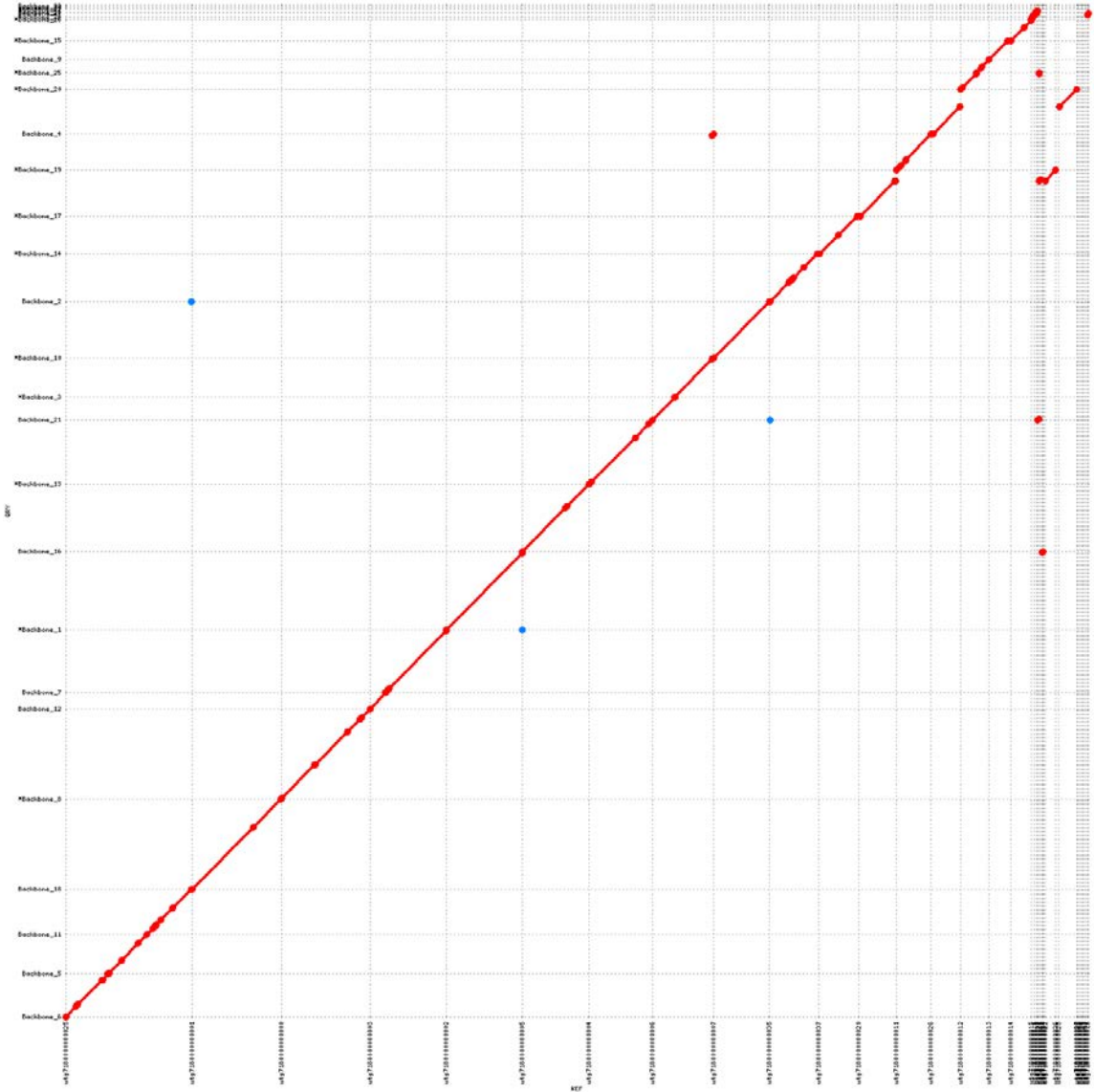

**Figure S9**. Alignment of high coverage PacBio assembly of *S. cer* genome (Y-axis) to the 454 assembly.

**Figure S10**.The alignment of DBG2OLC assembly of the *A. thaliana* genome (Y-axis) to the high coverage PacBio assembly (X-axis). 10x PacBio reads and 50x Illumina reads are used.

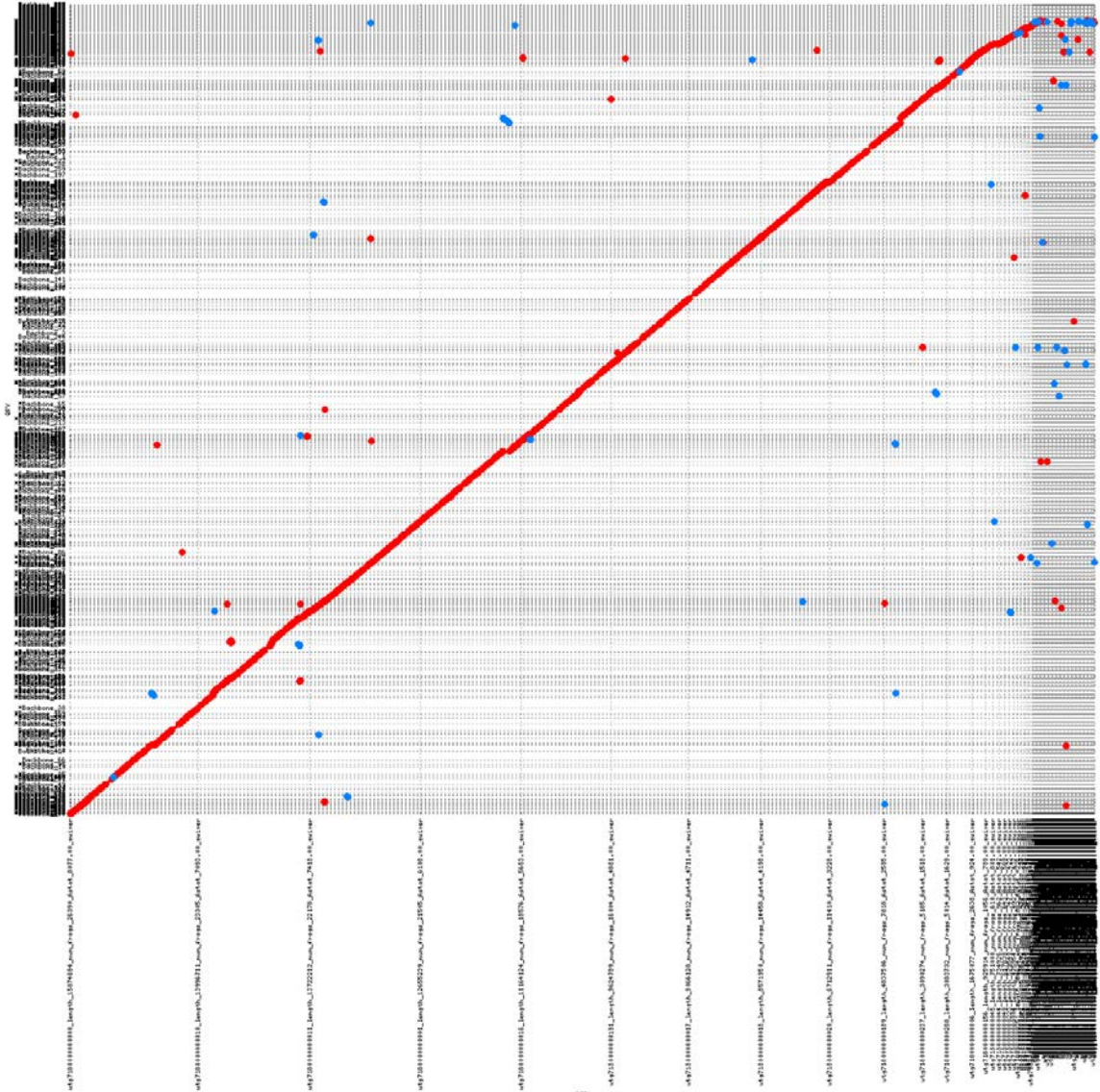

**Figure S11**.The alignment of DBG2OLC assembly of the *A. thaliana* genome (Y-axis) to the high coverage PacBio assembly (X-axis). 20x PacBio reads and 50x Illumina reads are used.

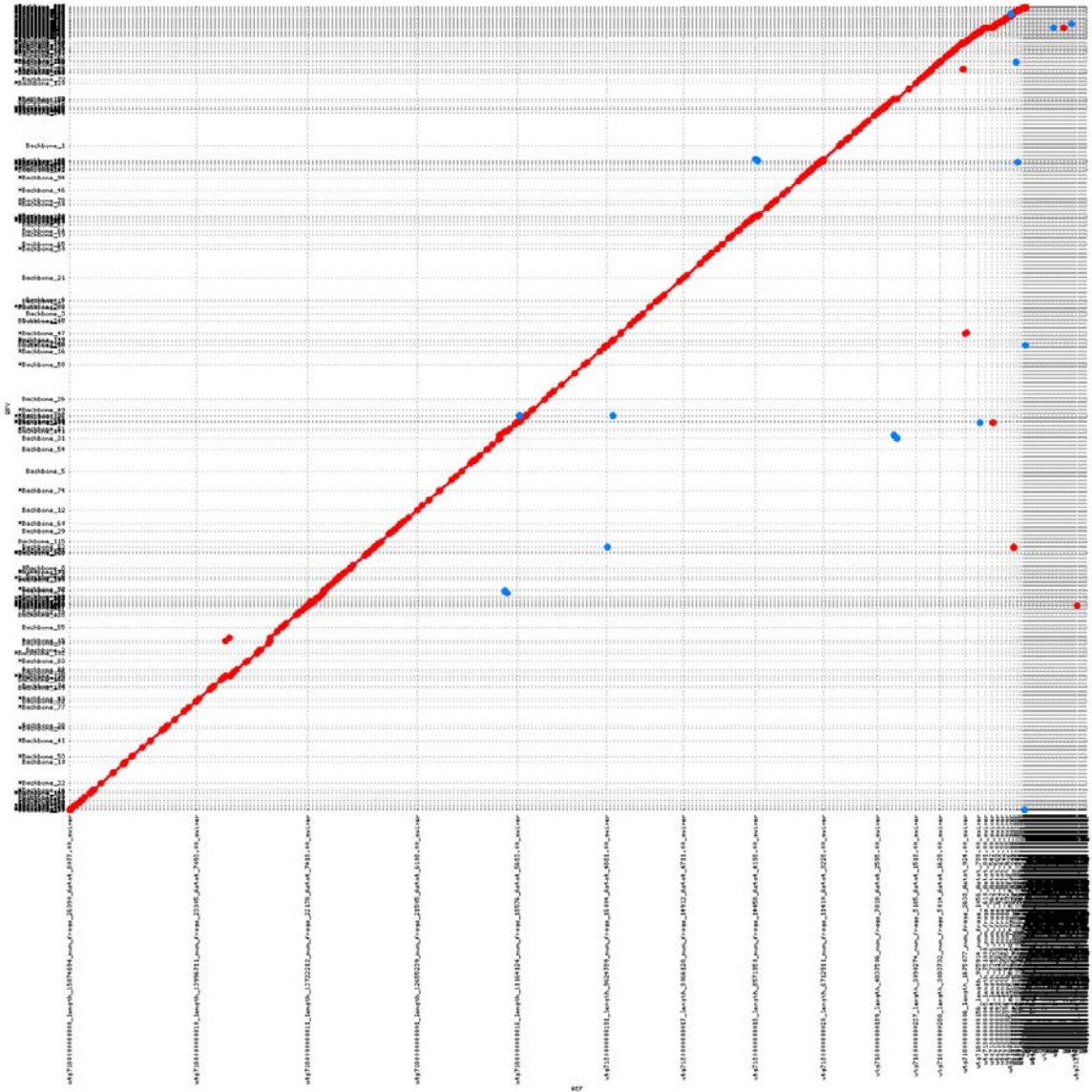

**Figure S12**.The alignment of DBG2OLC assembly of the *A. thaliana* genome (Y-axis) to the high coverage PacBio assembly (X-axis). 40x PacBio reads and 50x Illumina reads are used.

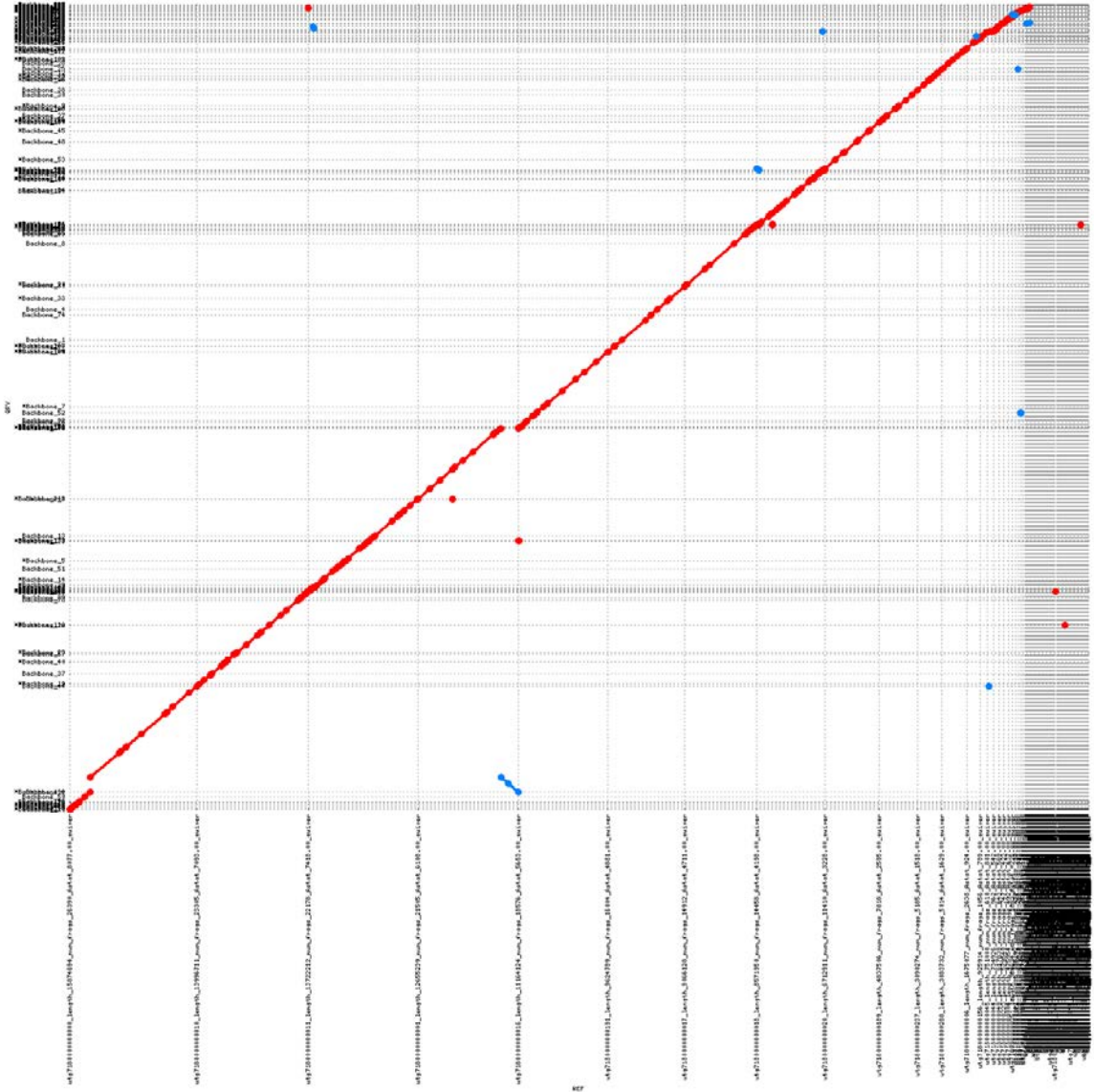

**Figure S13**. The alignment of DBG2OLC assembly of the *E. coli* genome (Y-axis) to the reference genome (X-axis). 30x Oxford Nanopore reads and 50x Illumina reads are used.

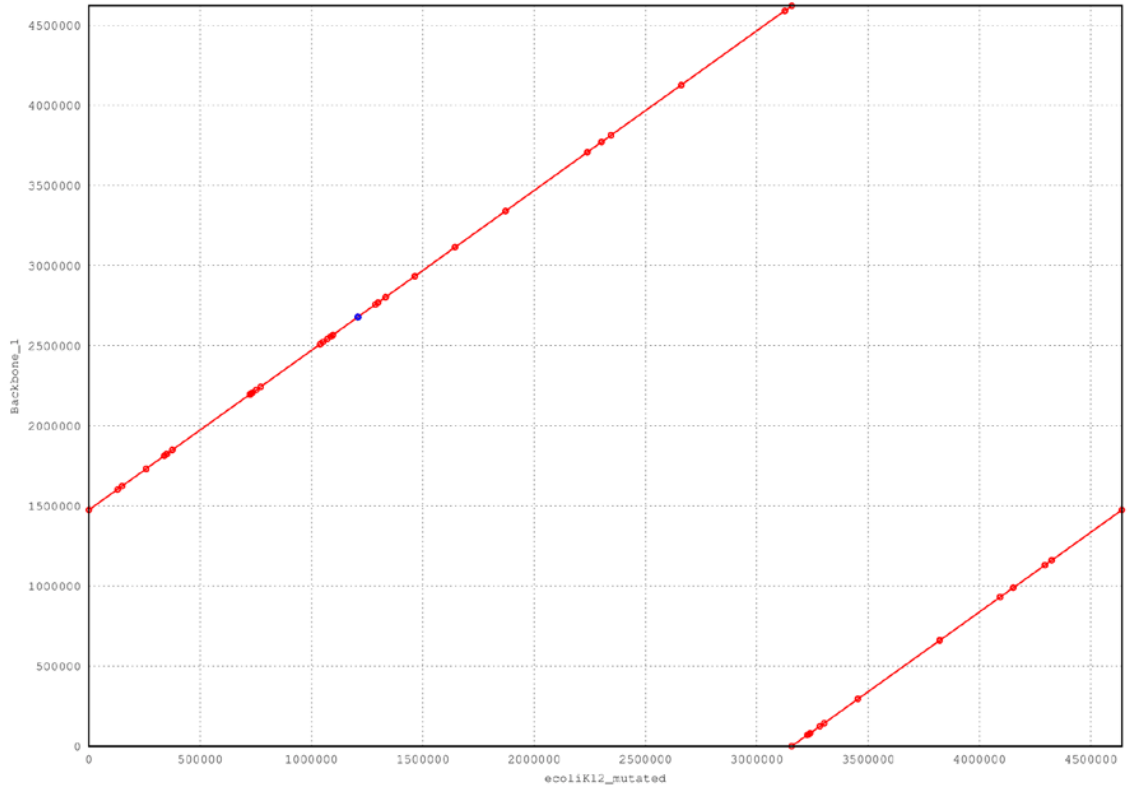

## 5. Commands used to run other pipelines

MHAP
PBcR -length 500 -partitions 200 -l yeast -s pacbio.spec -fastq 80x.fastq genomeSize=12000000

HGAP
smrtpipe.py -D NPROC=3 -D CLUSTER=BASH -D MAX_THREADS=4 --params=params.xml
        xml:input.xml >smrtpipe.log

CA
runCA -p yeast-trim -d yeast-trim -s yeast-trim.spec yeast-untrimmed.frg

PacBioToCA
pacBioToCA -length 500 -partitions 200 -l ec_pacbio -t 16 -s pacbio.spec -fastq 80x.fastq yeast.frg
runCA -p asm -d asm -s asm.spec ec_pacbio.frg

Falcon
fc_run.py fc_run.cfg

ECTools
for i in {0001..000N}; do cd $i; qsub -cwd -j y -t 1:500 ../correct.sh; cd ..; done
runCA -p assembly -d assembly -s assembly.spec organism.cor.frg

Celera assembler specification files can be found in a different attachment.

# References

1   Simpson, J. T. & Durbin, R. Efficient de novo assembly of large genomes using compressed data structures. *Genome research* **22**, 549-556, doi:10.1101/gr.126953.111 (2012).
2   Luo, R. *et al.* SOAPdenovo2: an empirically improved memory-efficient short-read de novo assembler. *GigaScience* **1**, 18, doi:10.1186/2047-217X-1-18 (2012).
3   Bankevich, A. *et al.* SPAdes: a new genome assembly algorithm and its applications to single-cell sequencing. *Journal of computational biology : a journal of computational molecular cell biology* **19**, 455-477, doi:10.1089/cmb.2012.0021 (2012).
4   Ferragina, P. & Manzini, G. in *Proceedings. 41st Annual Symposium on Foundations of Computer Science.*  390-398.
5   Ye, C., Ma, Z. S., Cannon, C. H., Pop, M. & Yu, D. W. Exploiting sparseness in de novo genome assembly. *BMC Bioinformatics* **13 Suppl 6**, S1, doi:10.1186/1471-2105-13-S6-S1 (2012).